\documentclass[12pt]{article}

\usepackage{amsmath}

\usepackage{feynmp}
\usepackage{thophys}

\newcommand{\Dl}{\overleftarrow{D}}
\newcommand{\Dr}{\overrightarrow{D}}
\newcommand{\dl}{\overleftarrow{\partial}}
\newcommand{\dr}{\overrightarrow{\partial}}
\newcommand{\vDl}{\overleftarrow{vD}}
\newcommand{\vDr}{\overrightarrow{vD}}

\newcommand{\Tr}[1]{\mathop{\mathrm{Tr}}\nolimits \left\{ #1 \right\}}

\newcommand{\vone}[1]{\fmfv{d.sh=circle,d.si=13mm,d.f=0,l=$A(iv\Dr)$,l.d=0}{#1}}
\newcommand{\vtwo}[1]{\fmfv{d.sh=circle,d.si=5mm,d.f=0,l=$B$,l.d=0}{#1}}
\newcommand{\vthree}[1]{\fmfv{d.sh=circle,d.si=7mm,d.f=0,l=$AB$,l.d=0}{#1}}
\newcommand{\lagone}[1]{\fmfv{d.sh=square,d.si=3.5mm,d.f=0}{#1}}
\newcommand{\lagtwo}[1]{\fmfv{d.sh=square,d.si=3.5mm,d.f=1}{#1}}
\newcommand{\curzero}[1]{\fmfv{d.sh=triangle,d.si=5.5mm,d.f=0}{#1}}
\newcommand{\curtwo}[1]{\fmfv{d.sh=triangle,d.si=5.5mm,d.f=1}{#1}}
\newcommand{\arr}{\fmfset{arrow_len}{1.5mm}\fmfset{arrow_ang}{25}}

\begin{document}
\allowdisplaybreaks
\title{%
  Heavy Quark Effective Field Theory at $\mathcal{O}(1/m_Q^2)$. II.\\
  QCD Corrections to the Currents}

\author{%
  Christopher Balzereit\thanks{%
      Supported by the German National Scholarship Foundation.}%
    \thanks{email:
      \texttt{chb@crunch.ikp.physik.th-darmstadt.de}} \\
  Thorsten Ohl\thanks{e-mail:
      \texttt{Thorsten.Ohl@Physik.TH-Darmstadt.de}} \\
  \hfil \\
  Technische Hochschule Darmstadt \\
  Schlo\ss gartenstr. 9 \\
  D-64289 Darmstadt \\
  Germany}

\date{%
  IKDA 96/12\\
  hep-ph/9612339\\
  December 1996}

\maketitle
\begin{abstract}
  We present a calculation of the renormalized heavy-light and
  heavy-heavy currents in HQET at order~$\mathcal{O}(1/m_Q^2)$.
\end{abstract}

\begin{fmffile}{pl2pics}
\section{Introduction}
\label{sec:introduction}

In~\cite{Balzereit/Ohl:1996:Lagrangian} we have presented a new
calculation of the~$\mathcal{O}(\alpha)$ renormalized Lagrangian of
Heavy Quark Effective Field Theory (HQET) at~$\mathcal{O}(1/m_Q^2)$.
This result has been confirmed using different calculational
strategies in~\cite{Finkemeier/McIrvin:1996:HQET} (see
also~\cite{Blok/etal:1996:spectator}).  In this note we proceed with
the results of a calculation of the renormalized heavy-light and
heavy-heavy currents in the same order.  The goal is a systematic
calculation of the matrix element
\begin{equation}
\label{eq:MEgen}
  \Braket{ D(v') | \bar c \gamma_\mu (1-\gamma_5) b | B(v) }
\end{equation}
up to and including~$\mathcal{O}(1/m_Q^2)$, including leading order
QCD radiative corrections.  

A full calculation of the matrix element~(\ref{eq:MEgen}) of the
heavy-heavy current at non-zero recoil is technically feasible, but
still of limited phenomenological value.  The appearance of the second
independent velocity vector~$v'$ leads to a proliferation of operators
whose matrix element would have to be parametrized.  This results in a
substantial decrease in predictive power. Instead we will
follow~\cite{Mannel:1994:zero_recoil} and concentrate on the zero
recoil case~$v=v'$, which should be considered as the lowest order of
an expansion of the matrix element (\ref{eq:MEgen}) at the symmetry
point~$\omega=vv'=1$.  In this case Luke's celebrated
theorem~\cite{Luke:1990:Theorem} guarantees that there are
no~$\mathcal{O}(1/m_Q)$ corrections to the current, but there are
non-trivial contributions
at~$\mathcal{O}(1/m_Q^2)$~\cite{Falk/Neubert:1993:higher_orders,
Mannel:1994:zero_recoil}.

This note is organized as follows: in section~\ref{sec:basis} we
define our operatorbasis for the effective lagrangian, heavy-light and
heavy-heavy currents up to and including~$\mathcal
O(1/m_{Q}^{2})$. Our result for the anomalous dimensions will be
presented in section \ref{sec:anodim} followed by a discussion of
various consistency checks in section~\ref{sec:checks}.  In
section~\ref{sec:RGE} the numerical solution of the renormalization
group equation (RGE) for the currents at~$\mathcal O(1/m_{Q^{2}})$ is
presented and is discussed with respect to phenomenological
applications.

\section{Operator basis}
\label{sec:basis}

\subsection{Heavy-light currents}
\label{sec:HL}

We start with the current up to~$\mathcal O(1/m_{Q})$ and the
operators in the effective lagrangian needed for the construction of
time-ordered products in~$\mathcal O(1/m^{2}_{Q})$.  The kinetic and
chromomagnetic operators in the effective lagrangian of dimensions 5
\begin{subequations}
\label{eq:lag12}
\begin{align}
   \mathcal{O}^{(1)c/b}_{1} &=
    \bar h^{c/b}_{v}(iD)^{2}h^{c/b}_{v}, &
   \mathcal O^{(1)c/b}_{2} &=
     \frac{g}{2}\bar h^{c/b}_{v} \sigma^{\mu \nu}G_{\mu \nu}h^{c/b}_{v} \\
\intertext{and~6}
   \mathcal O^{(2)c/b}_{1} &=
     \bar h^{c/b}_{v} iD_{\mu}(ivD)iD^{\mu}h^{c/b}_{v},&
   \mathcal O^{(2)c/b}_{2} &=
     \bar h^{c/b}_{v} i\sigma^{\mu\nu}iD_{\mu}(ivD)iD_{\nu}h^{c/b}_{v}
\end{align}
\end{subequations}
are labeled by a flavor index for future use.  We choose the
operatorbasis for the currents  of dimensions~3 and~4 as  
\begin{subequations}
\label{curr01}
\begin{align}
\label{eq:curr0}
  \mathcal O^{(0)}_{HL,1} &=
    \bar c_{L} v^{\mu} h_{v},&
  \mathcal O^{(0)}_{HL,2} &=
    \bar c_{L}\gamma^{\mu}h_{v} 
\end{align} 
and~$\mathcal O^{(1)}_{HL,i}$:
\begin{equation}
\label{eq:curr1}
 \begin{array}{cccc}
   -\bar c_{L} i \Dl^{\mu}h_{v}, &
    \bar c_{L} i \Dr^{\mu}h_{v}, &
   -\bar c_{L} (iv \Dl)v^{\mu}h_{v}, &
   -\bar c_{L} i \fmslash{\Dl}v^{\mu}h_{v}, \\
    \bar c_{L} i \fmslash{\Dr}v^{\mu}h_{v}, &
   -\bar c_{L} (iv\Dl) \gamma^{\mu}h_{v}, &
   -\bar c_{L} i \Dl_{\alpha} i \sigma^{\alpha \mu}h_{v}, &
    \bar c_{L} i \Dr_{\alpha} i \sigma^{\alpha \mu} h_{v}
\end{array}
\end{equation}
\end{subequations}
In~(\ref{eq:curr1}) the label~$i$ runs from left to right.  Because of
the momentum transfer at the weak current, we have to distinguish
derivatives~$\Dl=\dl +igT^{a}A^{a} $ acting to the left from
derivatives~$\Dr = \dr -igT^{a}A^{a} $ acting to the right in the
operator basis. It turns out that not all operators that contain two
of these derivatives are independent. In fact there are \emph{arrow
identities~(AI)}, that we have to take into account:
\begin{subequations}
\label{eq:AI}
\begin{align}
 -i\Dr_{\alpha}i\Dr_{\beta} +i\Dr_{\beta}i\Dr_{\alpha}
       -i\Dl_{\alpha}i\Dr_{\beta} +i\Dr_{\beta}i\Dl_{\alpha }=&0\\
 i\Dr_{\alpha}i\Dr_{\beta} -i\Dl_{\alpha}i\Dl_{\beta}
    -i\Dr_{\beta}i\Dr_{\alpha}+i\Dl_{\beta}i\Dl_{\alpha}=&0\\
 i\Dr_{\alpha}i\Dr_{\beta} +i\Dr_{\alpha}i\Dl_{\beta}
    -i\Dr_{\beta}i\Dr_{\alpha}-i\Dl_{\beta}i\Dr_{\alpha}=&0\\
\label{eq:arrowid}
 \bar c_L  \left[i\Dl_{\alpha}i\Dl_{\beta} +i\Dr_{\beta}i\Dr_{\alpha}
   +i\Dr_{\alpha}i\Dl_{\beta}  +i\Dr_{\beta}i\Dl_{\alpha} \right] h_v=&
 i\partial_{\alpha}i\partial_{\beta}\bar c_L h_v
\end{align}
\end{subequations}
The identities~$(\ref{eq:AI})$ are operator identities and we stress
that their derivation~\cite{Balzereit/Ohl:1996:technical} is
independent from the applicability of partial integration. The
identity~(\ref{eq:arrowid}) results in additional contributions to the
anomalous dimensions from the divergent parts of the two point
function with dimension 5 operator insertions.  The~AI and the
equation of motion~(EOM) for the heavy quark~$(ivD)h_v=0$ reduce the
full set of 56~local operators to 26~operators $\mathcal
O^{(2)}_{HL,i}$, which we choose as
\renewcommand{\arraystretch}{1.2}
\begin{equation}
  \begin{array}{ccccc}
      (i\vDl)^2 v^\mu,
    & i\Dl i\Dl  v^\mu,
    & -i\Dl i\Dr v^\mu,
    & i\Dr i\Dr v^\mu, \\
      -i\vDl i\fmslash\Dr v^\mu,
    & i\vDr i\fmslash\Dr v^\mu,
    & i\vDl i\fmslash\Dl v^\mu,
    & -i\Dl_\alpha i\Dr_\beta \sigma^{\alpha\beta} v^\mu, \\
      i\Dr_\alpha i\Dr_\beta \sigma^{\alpha\beta} v^\mu,
    & -i\vDl i\Dr^\mu,
    & i\vDr i\Dr^\mu,
    & i\Dl^\mu i\vDl, \\
      (i\vDl)^2 \gamma^\mu,
    & i\Dl i\Dl \gamma^\mu,
    & -i\Dl i\Dr \gamma^\mu,
    & i\Dr i\Dr \gamma^\mu, \\
      -i\fmslash\Dl i\Dr^\mu,
    & i\fmslash\Dr i\Dr^\mu,
    & -i\Dl^\mu i\fmslash\Dr,
    & i\Dr^\mu i\fmslash\Dr, \\
      i\Dl^\mu i\fmslash\Dl,
    & -i\vDl i\Dr_\alpha \sigma^{\alpha\mu},
    & i\vDr i\Dr_\alpha \sigma^{\alpha\mu},
    & i\vDl i\Dl_\alpha \sigma^{\alpha\mu}, \\
      -i\Dl_\alpha i\Dr_\beta v_\gamma i\epsilon^{\alpha\beta\gamma\mu},
    & i\Dr_\alpha i\Dr_\beta v_\gamma i\epsilon^{\alpha\beta\gamma\mu},
  \end{array}
\end{equation}
where the~$\bar c_L$ on the left and the~$h_v$ on the right have been
suppressed.  Since~$m_c \neq 0$, the full basis contains 18~operators
proportional to~$m_c$ as well as~$m_c^2$, which can be eliminated
using the EOM for the light quark
\begin{equation}
  m_c \bar c_R= -\bar c_L i\fmslash\Dl.
\end{equation}
In~\cite{Balzereit/Ohl:1996:Lagrangian} the operator basis could be
divided naturally in four classes: vanishing by the equations of
motion~(EOM) or not, local or time-ordered product.  This time we have
to consider another class of unphysical operators, that are related to
the other operators by~\emph{contraction identities~(CI)}.  These
identities arise in time-ordered products with $(ivD)$-operators
acting on \emph{internal} heavy quark
propagators~\cite{Falk/etal:1994:KI}. They are independent from the
external states and can be illustrated graphically as follows:
\begin{align*}
  \Tprod{[\ldots A(iv\Dr)h_{v}][\bar h_{v} B\ldots]}
     &\propto\Tprod{ \ldots AB \ldots} \\
  \parbox{45mm}{%
    \begin{fmfgraph*}(45,10)
      \fmfleft{hi}\fmfright{qo}
      \fmf{dbl_plain_arrow}{hi,op1}
      \fmf{dbl_plain_arrow}{op1,op2}
      \fmf{plain_arrow,tension=2}{op2,qo}
      \vone{op1}
      \vtwo{op2}
    \end{fmfgraph*}}
    & \propto
      \parbox{25mm}{
        \begin{fmfgraph*}(25,10)
          \fmfleft{hi}\fmfright{qo}
          \fmf{dbl_plain_arrow}{hi,op1}
          \fmf{plain_arrow}{op1,qo}
          \vthree{op1}
        \end{fmfgraph*}}
\end{align*}
Even though their derivation proceeds in terms of unrenormalized
operators, the~CI hold true under renormalization in the $MS$-scheme,
which we haved used.  The full basis contains 56~time-ordered products
of lower order contributions to the current with operators from the
effective lagrangian of the $b$-quark. After application of the~CI and
the~EOM for the heavy quark, we are left with 26~time-ordered
products. They are classified as follows:
\begin{itemize}
 \item double-insertions from the effective lagrangian in $\mathcal
    O(1/2m_{b})$:
   \begin{equation}
      \mathcal T^{(011)bb}_{HL,ijk}=-(1-\frac{1}{2}\delta_{jk})
      \Tprod{\mathcal O^{(0)}_{HL,i},\mathcal O^{(1)b}_{j},
        \mathcal O^{(1)b}_{k}}
      \quad i,j,k=1,2
   \end{equation}
   The prefactor accounts for insertions of identical operators.
   To avoid double-counting  
   the additional constraint~$j \leq k$ has to be imposed.
 \item single-insertions from the effective lagrangian in
   $\mathcal O(1/(2m_{b})^{2})$:
   \begin{equation}
     \mathcal T^{(02)b}_{HL,ij}=
     \Tprod{\mathcal O^{(0)}_{HL,i},\mathcal O^{(2)b}_{j}}
     \quad i,j = 1,2
   \end{equation}
 \item mixed insertions of dimension~4 currents and 
   the effective lagrangian in $\mathcal O(1/2m_{b})$:
   \begin{equation}
     \mathcal T^{(11)b}_{HL,ij}=
     \Tprod{\mathcal O^{(1)}_{HL,i},\mathcal O^{(1)b}_{j}}
     \quad 
     \begin{cases} i=1,\ldots,8\\ j = 1,2 \end{cases}
   \end{equation}
\end{itemize}

\subsection{Heavy-heavy currents at zero recoil}
\label{sec:HH}

The construction of the operator basis contributing to the heavy-heavy
current proceeds in analogy to the heavy-light current.  Since the
$c$-Quark is now static, the tensor structures of flavor changing
operators are projected from the left by~$P_v^+$ and we have to
consider additional time-ordered products with operators in the
effective lagrangian of the static $c$-Quark.  To lowest order only
two operators contribute:
\begin{align} 
  \bar h_v^c P_5^+ v^{\mu}h_v^b & &\bar h_v^c P_5^+ \gamma^{\mu}h_v^b
\end{align}
At order~$\mathcal O(1/m_Q)$, all operators sandwiched between
physical states vanish by Luke's theorem and may be discarded.  At
order~$\mathcal O(1/m^2_Q)$ the full basis contains 14~local and
100~nonlocal operators. After application of~CI and~EOM for the heavy
quarks, we are left with 46~operators.

The action of spin and flavor symmetry in the matrix elements of
time-ordered products reveals 20~additional relations.  Using flavor
symmetry, for example, the matrix elements of~$\mathcal
T^{(011)cb}_{HH,112}$ and~$\mathcal T^{(011)cb}_{HH,121}$ between
mesonic states~ $\vert \mathcal D \rangle$ and~$\vert \mathcal B
\rangle$ are described by the same reduced matrix element multiplied
by Clebsch-Gordan coefficients
\begin{align} 
  \Tr{\sigma_{\alpha\beta}\bar \mathcal
    M_{c}P_{5}^{+}v^{\mu}P_{v}^{+}\sigma^{\alpha\beta}\mathcal M_{b}},&&
  \Tr{\sigma_{\alpha\beta}\bar \mathcal
    M_{c}\sigma^{\alpha\beta}P_{v}^{+}P_{5}^{+}v^{\mu}\mathcal M_{b}}
\end{align}
with~$P_{5}^{+}v^{\mu}$, $\sigma^{\alpha\beta}$ and~$\mathcal M_{c/b}$
representing the Dirac-structure of the current, the chromomagnetic
operator and the external states respectively.  Since the former is
sandwiched between projectors~$P_v^+$, it conserves spin symmetry and
may be moved around freely inside the traces.  Consequently the traces
are equal and so are the matrix elements.  Other relations among
matrix elements can be established analoguously.

Furthermore, since spin symmetry allows us to rewrite all mixed
time-ordered products in terms of EOM-operators, they are removed from
the basis applying the~CI.

The basis is reduced by these relations to 6~local operators
\begin{align}
     (iD)^{2}v^{\mu},&
    &(iD)^{2}\gamma^{\mu},&
    &i\fmslash{D}iD^{\mu},&
    &iD^{\mu}i\fmslash{D},&
    &iD_{\alpha}iD_{\beta}i\sigma^{\alpha \beta}v^{\mu},&
    &iD_{\alpha}iD_{\beta}v_{\gamma}i\epsilon^{\alpha \beta \gamma \mu}
\end{align}
taken between static fields~$\bar h_{v}^{c}P_{5}^{+}$ and~$h_{v}^{b}$
and 20~time-ordered products
\begin{align}
   \mathcal T^{(011)bb}_{HH,ijk}\vert_{j\leq k},&
  &\mathcal T^{(011)cc}_{HH,2j2},&
  &\mathcal T^{(011)cb}_{HH,ijk}\vert_{j\leq k~\text{for}~ i = 1},&
  &\mathcal T^{(02)b}_{HH,ij},&
  &\mathcal T^{(02)c}_{HH,22},
\end{align}
where all indices run from~1 to~2.

\section{Anomalous dimensions}
\label{sec:anodim}

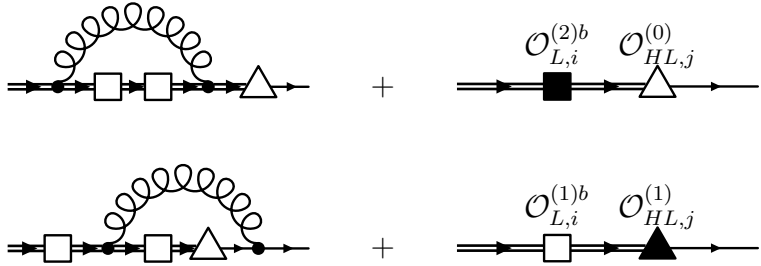
\begin{figure}
  \begin{align*}
    \parbox{40mm}{%
      \begin{fmfgraph}(40,20)\arr
        \fmfleft{hi}\fmfright{lo}
        \fmf{dbl_plain_arrow}{hi,v1,op1,op2,v2,cur}
        \fmf{plain_arrow}{cur,lo}
        \fmffreeze
        \fmf{gluon,left,tension=0}{v1,v2}
        \curzero{cur}
        \lagone{op1}
        \lagone{op2}
        \fmfdot{v1,v2}
      \end{fmfgraph}}\quad\quad+&\quad\quad
    \parbox{40mm}{%
      \begin{fmfgraph*}(40,20)\arr
        \fmfleft{hi}\fmfright{lo}
        \fmf{dbl_plain_arrow}{hi,op1,cur}
        \fmf{plain_arrow}{cur,lo}
        \fmfv{l.a=90,l.d=2.5mm,l=$\mathcal O_{L,,i}^{(2)b}$}{op1}
        \fmfv{l.a=90,l.d=2.5mm,l=$\mathcal O_{HL,,j}^{(0)}$}{cur}
        \curzero{cur}
        \lagtwo{op1}
      \end{fmfgraph*}}\\
    \parbox{40mm}{%
      \begin{fmfgraph}(40,20)\arr
        \fmfleft{hi}\fmfright{lo}
        \fmf{dbl_plain_arrow}{hi,op1,v1,op2,cur}
        \fmf{plain_arrow}{cur,v2,lo}
        \fmffreeze
        \fmf{gluon,left,tension=0}{v1,v2}
        \curzero{cur}
        \lagone{op1}
        \lagone{op2}
        \fmfdot{v1,v2}
      \end{fmfgraph}}\quad\quad+&\quad\quad
    \parbox{40mm}{%
      \begin{fmfgraph*}(40,20)\arr
        \fmfleft{hi}\fmfright{lo}
        \fmf{dbl_plain_arrow}{hi,op1,cur}
        \fmf{plain_arrow}{cur,lo}
        \fmfv{l.a=90,l.d=2.5mm,l=$\mathcal O_{L,,i}^{(1)b}$}{op1}
        \fmfv{l.a=90,l.d=2.5mm,l=$\mathcal O_{HL,,j}^{(1)}$}{cur}
        \curtwo{cur}
        \lagone{op1}
      \end{fmfgraph*}}
  \end{align*}
  \caption{\label{fig:diagillust}%
     Nonlocal contributions to the renormalization
     of~$\vec{\mathcal{T}}^{(011)bb}_{HL}$. The operators of the
     effective lagrangian in~$\mathcal O(1/m_b)$ and 
     the current in~$\mathcal O(1/m^0_b)$ are represented by open
     squares and triangles respectively, whereas filled symbols
     represent the corresponding operators of the next dimension.}
\end{figure}

As in~\cite{Balzereit/Ohl:1996:Lagrangian}, we have calculated the
anomalous dimensions in the background field gauge with gauge
parameter~$\xi$.  For the physical operator bases presented here, all
$\xi$-dependence has to drop out, which provides a powerful
cross-check.  In the case of heavy-light currents, the anomalous
dimensions in leading and subleading order are well
known~\cite{Falk/etal:1990:HQET_HL,Neubert:1994:HQET_review,
Kilian:1994:thesis}
and are reproduced by our calculations.  Our result in~$\mathcal
O(1/m^{2}_{b})$ can be written in block form, separating local
operators from time-ordered products:
\begin{equation}
\label{eq:anodim2HL}
  \hat\gamma^{(2)}_{HL} =
    \begin{matrix}
      \vphantom{{\displaystyle\int}}
        & \begin{matrix}
              \vec{\mathcal{O}}^{(2)}_{HL}
            & \vec{\mathcal{T}}^{(11)b}_{HL}
            & \vec{\mathcal{T}}^{(02)b}_{HL}
            & \vec{\mathcal{T}}^{(011)bb}_{HL}
          \end{matrix} \\
      \begin{matrix}
        \vec{\mathcal{O}}^{(2)}_{HL}\\
        \vec{\mathcal{T}}^{(11)b}_{HL}\\
        \vec{\mathcal{T}}^{(02)b}_{HL}\\
        \vec{\mathcal{T}}^{(011)b}_{HL}
      \end{matrix} &
      \begin{pmatrix}
        \hat{\gamma}^{(2)}_{d} & 0 & 0 & 0\\
        \hat{\gamma}^{(11)}_{n}&\hat{\gamma}^{(11)}_{d}&0&0\\
        \hat{\gamma}^{(02)}_{n}&0&\hat{\gamma}^{(02)}_{d}&0\\
        \hat{\gamma}^{(011)}_{n,1}&\hat{\gamma}^{(011)}_{n,2}
         &\hat{\gamma}^{(011)}_{n,3}
         &\hat{\gamma}^{(011)}_{d}
      \end{pmatrix}
    \end{matrix} \,.
\end{equation}
Weinberg's theorem guarantees that local operators need no non-local
counterterms. This accounts for the vanishing of seven submatrices
in~(\ref{eq:anodim2HL}). There are, however, nonlocal
contributions~$\hat{\gamma}^{(011)}_{n,2}$
and~$\hat{\gamma}^{(011)}_{n,3}$ that arise from the renormalization
of the lagrangian or the lower order current.  This situation is
depicted diagramatically in figure~\ref{fig:diagillust}.

Our result for the heavy-heavy currents at the same order 
is of similiar structure:
\begin{equation}
\label{eq:anodim2HH}
  \hat\gamma^{(2)}_{HH} =
    \begin{matrix}
      \vphantom{{\displaystyle\int}}
        & \begin{matrix}
              \vec{\mathcal{O}}^{(2)}_{HH}
            & \vec{\mathcal{T}}^{(02)}_{HH}
            & \vec{\mathcal{T}}^{(011)}_{HH}
          \end{matrix} \\
      \begin{matrix}
        \vec{\mathcal{O}}^{(2)}_{HH}\\
        \vec{\mathcal{T}}^{(02)}_{HH}\\
        \vec{\mathcal{T}}^{(011)}_{HH}
      \end{matrix} &
      \begin{pmatrix}
        \hat{\gamma}^{(2)}_{d} & 0 & 0\\
        \hat{\gamma}^{(02)}_{n}&\hat{\gamma}^{(02)}_{d}&0\\
        \hat{\gamma}^{(011)}_{n,1}
          &\hat{\gamma}^{(011)}_{n,2}&\hat{\gamma}^{(011)}_{d}
      \end{pmatrix}
    \end{matrix} \,.
\end{equation}
The entries for all submatrices~$\hat\gamma$ are available from the
authors on request.  A typical example is given by the anomalous
dimensions of the local operators:
\begin{equation}
  \hat{\gamma}^{(2)}_{HH,d} = \frac{C_{A}}{4} \cdot
     \begin{pmatrix}
       0&0&0&0&0&0\\
       0&0&0&0&0&0\\
       0&0&-1&1&0&0\\
       0&0&1&-1&0&0\\
       0&0&0&0&-2&0\\
       0&0&0&0&0&-2
     \end{pmatrix}.
\end{equation}
The observation, that the kinetic and chromomagnetic operators (the
upper left and the lower right $2\times 2$-block) renormalize as their
counterparts in the effective lagrangian in~$\mathcal O(1/m_Q)$, is
easily explained by the fact that both operator bases coincide at zero
recoil up to parity violating contributions.

The anomalous dimensions have been calculated with the help of
\texttt{FORM}~\cite{Vermaseren:1991:FORM2}.  In the background field
gauge, the divergent part of the three-point function has to be
calculated to derive the anomalous dimensions.  The latter receive
additional contributions from the two-point function only in the case
of heavy-light currents as a result of the
identity~(\ref{eq:arrowid}).

\section{Consistency checks}
\label{sec:checks}

In section~\ref{sec:basis} we have used several identities to reduce
our operatorbasis.  However, we have to verify that this reduction is
compatible with the~RGE.  Suppose that a linear combination
\begin{equation}
\label{eq:red}
  O_\kappa = \sum_i \kappa_i O_i
\end{equation}
of operators does not contribute to the matrix element under
consideration
\begin{equation}
  \Braket{ D(v) | O_\kappa | B(v) } = 0
\end{equation}
In general, this relation will not be stable under renormalization since
\begin{equation}
  O_{\delta\kappa} = \sum_i (\delta\kappa)_i O_i
    = \sum_i \kappa_i \frac{d}{d\ln\mu} O_i
    = \sum_{i,j} \kappa_i \gamma_{ij} O_j
\end{equation}
and
\begin{equation}
  \Braket{ D(v) | O_{\delta\kappa} | B(v) } \not= 0
\end{equation}
unless~$\gamma_{ij}$ satisfies further conditions.

For a consistent reduction of the operatorbasis we have to start with
the anomalous dimensions of the full operatorbasis.  Whenever an
operator identity~(\ref{eq:red}), e.g.~CI or EOM, is applied to reduce
the full operatorbasis, we have verified its compatibility with the
RGE.  Also, the relations between matrix elements of heavy-heavy
currents have been shown stable under renormalization.

In the background field gauge, only the anomalous dimensions of
operators that are manifestly invariant under gauge transformations of
the background field have to be calculated.  The Ward identities of
the underlying BRST-symmetry are particularly simple and provide a
powerful cross-check of our
results~\cite{Balzereit/Ohl:1996:Lagrangian}. From the Ward identities
for the background field and the gauge parameter dependence of the
effective action follows, that the counterterms proportional to~$C_A$
have to be transversal and should not depend on~$\xi$. Our results
pass these consistency checks.

Additional checks are provided by the construction of relations among
operators. For example, the differentiation of the lowest order
currents
\begin{equation}
       \label{eq:konspart}
       i\partial_{\nu}\bar c_{L} \Gamma^{\mu} h_{v}  = 
        \bar c_{L}i\Dl_{\nu} \Gamma^{\mu} h_{v}
       + \bar c_{L} \Gamma^{\mu} i\Dr_{\nu}h_{v}
\end{equation}
leaves their anomalous dimensions unaffected and relates them to
subleading operators.  The operators on the right hand side must mix
under renormalization in such a way, that their sum can be expressed
by the total derivative of the current on the left hand side, which
renormalizes multiplicatively.  Our result is consistent with all
relations of this type, including generalizations to local operators
of higher dimension and time-ordered products.

\section{Solving the renormalization group equations}
\label{sec:RGE}
\subsection{Matching}
As in any effective field theory calculation, we sum the radiative
corrections by matching the operators at the thresholds and by running
their  coefficients between thresholds.  In the case at hand, we have two
thresholds~$\mu=m_b$ and~$\mu=m_c$.  It is convenient to express the
matching conditions as 
matrices~$\mathcal M^b$ 
\begin{equation}
  \Braket{ c | \bar c \gamma_\mu (1-\gamma_5) b | b }_{\mu=m_b}
    = \sum_i \mathcal M^b_{i}
        \Braket{ c | \mathcal{O}_{HL,i} | b(v) }_{\mu=m_b}
\end{equation}
and~$\mathcal M^c$
\begin{equation}
   \Braket{ c | \mathcal{O}_{HL,i} | b(v) }_{\mu=m_c}
    =\sum_j \mathcal M^c_{ij}
        \Braket{ c(v) |  \mathcal{O}_{HH,j} | b(v) }_{\mu=m_c}\,.
\end{equation}
Since there is only one QCD-operator, $\mathcal M^b$ reduces to a
vector.  The $\mathcal{O}_{HL/HH,j}$ contain all physical local
operators and time-ordered products up to and including dimension 5,
that contribute to heavy-light and heavy-heavy currents respectively.
Consequently the matching matrices receive contributions suppressed by
inverse powers of the c- and b-quark masses up to and
including~$\mathcal O(1/m_{c/b}^2)$.  In addition, the matching of
HL-operators with a covariant derivative acting on~$\bar c$ induces a
nontrivial $m_c/m_b$-dependence in~$\mathcal M^c$.  For example, we
have to lowest order
\begin{equation}
   -\bar c i\Dl_{\mu} \rightarrow -\bar h^c_v e^{im_cvx}i\Dl_{\mu} 
       = -\bar h^c_v i\Dl_{\mu}e^{im_cvx}
          + m_c \bar h^c_v v_{\mu} e^{im_cvx}.
\end{equation}

\subsection{RG-Running}

For the lagrangian at~$\mathcal O(1/m_Q^2)$, only five operators do
not vanish by the EOM~\cite{Balzereit/Ohl:1996:Lagrangian} and an
analytical solution of the RGE is straightforward.  In contrast, the
operator bases for the heavy-light and heavy-heavy currents contain~52
and~26 operators respectively and an analytical solution of the~RGE is
impractical, to say the least

In each order~$1/m_Q$, the solution of the~RGE is expressed in terms
of evolution matrices
\begin{equation}
\label{eq:RHL}
  R^{(n)}_{\text{HL/HH}}(\mu,m_{c/b})
    = \exp \left( - \frac{1}{2\beta^{(1)}}
                    \ln\left(\frac{\alpha(\mu)}{\alpha(m_{c/b})}\right)
                    \cdot \hat\gamma^{(n)T}_{\text{HL/HH}}\right).
\end{equation}
In~$\beta^{(1)}=(33-2n_f)/12$, the number of active flavors takes the
values~$n_f=4$ and~$n_f=3$ for heavy-light and heavy-heavy currents
respectively.

Introducing their direct sum
\begin{align}
   \mathcal R_{HL} &=
      R^{(0)}_{\text{HL}} \oplus R^{(1)}_{\text{HL}}
        \oplus R^{(2)}_{\text{HL}}\\ 
\label{eq:evMHH}
   \mathcal R_{HH} &=
      R^{(0)}_{\text{HH}} \oplus R^{(1)}_{\text{HH}}
        \oplus R^{(2)}_{\text{HH}}\,,
\end{align}
the combined solution for the Wilson coefficients at the hadronic
scale can be written in compact form:
\begin{equation}
\label{eq:combsol}
  C_{\text{HH}}(\Lambda_{\text{hadr}})
    = \mathcal R_{\text{HH}}(\Lambda_{\text{hadr}},m_c)
       \cdot \mathcal M^c \cdot \mathcal R_{HL}(m_c,m_b)
       \cdot \mathcal M^b \,.
\end{equation}
The lowest order currents are conserved at zero recoil,
therefore~$R^{(0)}_{\text{HH}}=1$ in~(\ref{eq:evMHH}).
Furthermore, we can set~$R^{(1)}_{\text{HH}}=0$ because Luke's theorem
renders the corresponding contribution irrelevant.

For practical purposes, it is most convenient to sum the power series
of the exponential in~(\ref{eq:RHL}) directly.  This series converges
fast and any desired accuracy can be obtained.  This procedure is used
in the numerical results in table~\ref{tab:HH3}, where a factor
of~$1/(2m_c)^2$ must be attached to the higher order coefficients. We
also present the tree-level values of the coefficients to expose the
effect of the radiative corrections.
 
We stress that the matrices~$\hat\gamma^{(2)}_{HL/HH}$ are defective,
i.e.~do not have a complete set of eigenvalues. Therefore the analysis
of~\cite{Ohl/etal:1993:DDbar} is not applicable and we can not derive
general properties of the~RG evolution from the spectrum of the
matrices.

\subsection{Results}

Let us finally discuss phenomenological applications of our result.

{}From the coefficient~$C^{(0)}_{HH,2}$ of the lowest order symmetry
breaking operator in the top row of table~\ref{tab:HH3}, one obtains a
value of~1.180 for the short-distance coefficient~$\eta_{A}$ of the
axial vector current which is larger than~$1$.  This contradicts the
established result in~\cite{Neubert:1994:HQET_review} which is smaller
than~$1$.  This apparent contradiction is resolved if one takes into
account the complementarity of the calculational approaches.

Our calculation uses effective field theory to resum
corrections~$\alpha_s \ln m_b/m_c$ from the running between~$m_b$
and~$m_c$, 
while~\cite{Neubert:1994:HQET_review} integrates out the 
b- and c-quarks simultanuously.
The latter approach resums the powers of~$z=m_c/m_b$.
After applying radiative corrections, \cite{Neubert:1994:HQET_review}
includes~$\alpha_s^n \ln^{n-1}z$, $\alpha_s^n \ln^{n}z\cdot z$
and $\alpha_s \ln z\cdot z^2$, while we have concentrated
on~$\alpha_s^n \ln^n z$, $\alpha_s^n \ln^n z\cdot z$
and~$\alpha_s^n \ln^n z\cdot z^2$.
Therefore our result for~$\eta_A$ should not be taken face value.
Instead, our terms~ $\alpha_s^n \ln^n z\cdot z^2$ with~$n>1$ are 
corrections to the result of~\cite{Neubert:1994:HQET_review}.
In practice this means that the term~$\alpha_s \ln z \cdot z^2$ should
be subtracted from our result an the remaining correction should be
added to the result of~\cite{Neubert:1994:HQET_review}.  This
correction is~$0.002$ and the final result for~$\eta_A$ remains
below~$1$.

Up to~$\mathcal O(1/m_{c/b})$ no form factors are needed to
parameterize the matrix elements in the effective theory. The lowest
order is normalized and Luke's theorem states the absence of
contributions at~$\mathcal O(1/m_{c/b})$.
 
In~$\mathcal O(1/m_{c/b}^2)$ the constraints of spin and flavor
symmetry are less powerful.  An expansion of the matrixelements in
terms of 12~nonperturbative formfactors has been performed
in~\cite{Mannel:1994:zero_recoil} on tree-level.

Our result can be used to extend this analysis to include
short-distance corrections.  However, Lukes's theorem, which has been
used extensively in~\cite{Mannel:1994:zero_recoil}, receives radiative
corrections at~$\mathcal O(1/m_{c/b}^2)$. Therefore the analysis
of~\cite{Mannel:1994:zero_recoil} has to be refined to include these
corrections.

\begin{table}
  \begin{center}
    \begin{tabular}{|c|r|r||c|r|r|} \hline
        &$\alpha^{0}$ &$\mu=\Lambda_{hadr}$&
        &$\alpha^{0}$ &$\mu=\Lambda_{hadr}$ \\ 
      \hline\hline
      $C^{(0)}_{HH,1}$&$0.000$&$-0.005$  &
      $C^{(0)}_{HH,2}$&$1.000$&$1.180$  \\ \hline
      $C^{(2)}_{HH,2}$&$-0.667$&$-0.534$  &
      $C^{(2)}_{HH,6}$&$-1.444$&$-2.596$\\
      $C^{(2)}_{HH,7}$&$-1.222$&$-1.702$  &
      $C^{(2)}_{HH,8}$&$0.556$&$1.001$  \\
      $C^{(2)}_{HH,13}$&$2.111$&$1.825$  &
      $C^{(2)}_{HH,14}$&$-1.444$&$-1.832$ \\ \hline
      $C^{(02)b}_{HH,11}$&$0.000$&$0.011$ &
      $C^{(02)b}_{HH,21}$&$-1.111$&$-2.747$ \\
      $C^{(02)b}_{HH,12}$&$0.000$&$-0.002$&
      $C^{(02)b}_{HH,22}$&$0.111$&$0.023$ \\
      $C^{(02)c}_{HH,22}$&$1.000$&$0.421$ &
      $C^{(011)bb}_{HH,111}$&$0.000$&$-0.005$ \\
      $C^{(011)bb}_{HH,211}$&$1.111$&$1.302$ &
      $C^{(011)bb}_{HH,112}$&$0.000$&$-0.004$ \\
      $C^{(011)bb}_{HH,212}$&$0.111$&$0.072$ &
      $C^{(011)bb}_{HH,122}$&$0.000$&$-0.002$ \\
      $C^{(011)bb}_{HH,222}$&$0.111$&$0.043$ &
      $C^{(011)cc}_{HH,212}$&$1.000$&$0.800$ \\
      $C^{(011)cc}_{HH,222}$&$1.000$&$0.543$ &
      $C^{(011)cb}_{HH,111}$&$0.000$&$-0.001$\\
      $C^{(011)cb}_{HH,211}$&$0.333$&$0.386$&
      $C^{(011)cb}_{HH,112}$&$0.000$&$-0.001$ \\
      $C^{(011)cb}_{HH,212}$&$0.333$&$0.229$&
      $C^{(011)cb}_{HH,221}$&$0.333$&$0.262$ \\
      $C^{(011)cb}_{HH,122}$&$0.000$&$0.000$ &
      $C^{(011)cb}_{HH,222}$&$0.333$&$0.155$ \\ \hline
    \end{tabular}
  \end{center}
  \caption{\label{tab:HH3}%
    Wilson coefficients in tree-level and one-loop
    approximation. A factor~$1/(2m_{c})^{2}$ is to be attached 
    to the coefficients of higher order operators.
    For the quark masses we use the values~$m_b=4.5 \mathrm{GeV}$
    and~$m_c=1.5 \mathrm{GeV}$ and for the  initial value of the
    running coupling constant  we use~$\alpha_s(M_Z)=0.120$.}
 \end{table} 
\section{Conclusions}
\label{sec:conclusions}

We have presented a calculation of renormalized heavy-light and
heavy-heavy currents at order~$\mathcal O(1/m_Q^2)$.  The anomalous
dimensions of the relevant operatorbasis of dimension~5 pass various
consistency checks.  We have solved the renormalization group equation
numerically for the Wilson-coefficients of heavy-light and heavy-heavy
currents, matching both quantities at the $m_{c}$-threshold.  We have
presented the numerical values of the coefficients of the heavy-heavy
current at the hadronic scale.

The two expansion parameters~$\omega-1$ and~$\Lambda_{hadr}/m_Q$ are
numerically of the same order (for currently accessible values
of~$\omega$). Therefore phenomenological applications should consider
also pieces of~$\mathcal O((\omega -1)^2)$ as well as~$\mathcal
O((\omega -1)\Lambda_{hadr}/m_Q)$.  While the former have been
available for a long time~\cite{Falk/etal:1990:HQET_trace}, the latter
have been inferred by reparametrization invariance arguments, which
should be used with care~\cite{Balzereit/Ohl:1996:Lagrangian}. A
direct calculation will be the subject of future investigations.
 
A more detailed discussion of the consistency checks provided by BRST
and reparameterization invariance will be presented
elsewhere~\cite{Balzereit/Ohl:1996:technical}, together with technical
details of the calculation of renormalized Lagrangian and currents.


\end{fmffile}
\end{document}